\documentclass[english,aps,prb,superscriptaddress,reprint]{revtex4-2}

\usepackage{graphicx}  
\usepackage{float}     
\usepackage{dcolumn}   
\usepackage{bm}        
\usepackage{amssymb}   
\usepackage{amsmath}
\usepackage{lipsum}    
\usepackage{xcolor}
\usepackage{soul}
\usepackage{float}
\usepackage{placeins}
\usepackage{physics}
\usepackage{mathrsfs}
\usepackage{bbold}
\usepackage{hyperref}
\hypersetup{
    colorlinks=true,     
    linkcolor=blue,      
    citecolor=blue,      
    urlcolor=blue,       
    }

\usepackage[utf8]{inputenc}
\usepackage{natbib}
\usepackage{graphicx}
\usepackage{subcaption}
\usepackage{booktabs}
\usepackage{blindtext}

\usepackage{color}

\definecolor{goodred}{RGB}{183,15,58}
\definecolor{goodblue}{RGB}{93,128,180}

\begin{document}


\title{Ultrafast switching of antiferromagnetic order by field-derivative torque}
\author{Pratyay Mukherjee}
\email[]{23dr0111@iitism.ac.in}
\affiliation{Department of Physics, Indian Institute of Technology (Indian School of Mines) Dhanbad, IN-826004, Dhanbad, India}

\author{Ritwik Mondal}
\email[]{ritwik@iitism.ac.in}
\affiliation{Department of Physics, Indian Institute of Technology (Indian School of Mines) Dhanbad, IN-826004, Dhanbad, India}

\begin{abstract}
Control of magnetic order in antiferromagnets is a central challenge in the development of next-generation spintronic devices. Here, we propose and analyze magnetization switching driven by the field-derivative torque, a torque that originates from the time-derivative of an applied THz pulse acting on the staggered order parameter. Using atomistic spin simulations, we show that the field-derivative torque couples efficiently to the Néel vector, enabling deterministic switching without net spin accumulation. Further, we show that using the circularly polarised THz pulse, the FDT-induced magnetization switching reduces the required THz magnetic field by two-fold. To this end, we compute the switching and non-switching areas as a function of THz pulse width, THz magnetic field, and damping of the antiferromagnetic material. We find that the switching and non-switching areas are completely deterministic in antiferromagnets. Moreover, the switching area increases by about 55\% when the FDT is considered.      
\end{abstract}

\maketitle

\section*{Introduction}
The ever-growing volume of data generated by modern technologies necessitates ultrafast data processing and high-speed storage devices. Conventional magnetic memory technologies, based on ferromagnetic materials, are limited by magnetization reversal timescales on the order of nanoseconds. In contrast, antiferromagnetic (AFM) materials offer a promising pathway toward faster magnetic memory owing to their intrinsic terahertz (THz) spin dynamics \cite{Lu2018APL,shalaby2018,donges2017}. With resonance frequencies in the THz regime, AFM can achieve magnetization switching on a picosecond timescale. Recent theoretical and experimental studies have demonstrated all-optical switching of AFM order using ultrafast THz pulses \cite{Mangin2014,Radu2011,vahaplar12,Ross2024,Dannerger2021PRBL}. In particular, experiments employing circularly polarized light have shown that such switching can occur through both thermal and nonthermal mechanisms, establishing AFM as a key platform for next-generation THz spintronic devices \cite{Wienholdt2013,Nishitani2012PRB,KIMEL20201,Kimel2005,Torre2021RMP,Kimel2009,Kampfrath2011,BaierlPRL2016,Baierl2016NonlinearSpinControl}.

{Magnetization dynamics in such experiments are typically governed by the stochastic Landau-Lifshitz-Gilbert (LLG) equation~\cite{ostler12,Kazantseva2008,Kazantseva2007,Atxitia2009,Chimata2015,Evans2014,Evans2012,florian2022,nowak2007handbook,John2017,Moreno2019PRB,MorenoPRB2017,Skubic_2008}.
This dynamical equation describes the time evolution of the magnetization, incorporating two torque contributions: one governing the precession of the magnetic moment and the other accounting for the damping of that precession \cite{landau35,Gilbert2004}.  Theoretical derivations of the LLG equation that include relativistic spin-orbit coupling lead to an additional spin torque term known as the field-derivative torque 
(FDT)~\cite{Mondal2016,Mondal2018PRB}. It was proposed that such a spin torque term has to be accounted for in a high spin-orbit coupling material driven by a THz pulse \cite{Mondal2019PRB,Blank2021THz}. We note that the LLG equation has also been augmented to include several nonrelativistic and relativistic spin torques, e.g., spin transfer torque~\cite{Ralph2008,slonczewski96,Berger1996}, spin-orbit torque~\cite{Gambardella2011,Manchon2019RMP}, optical spin-orbit torque~\cite{tesarova13,Huang2024,mondal2021terahertz}, and inertial spin torque~\cite{Ciornei2011,neeraj2019experimental,unikandanunni2021inertial,Mondal2017Nutation,De2025experiment,Li2022APL}. }  

Materials with strong spin–orbit coupling typically exhibit enhanced Gilbert damping. When such materials are exposed to ultrafast THz magnetic fields, the FDT acts in conjunction with the conventional Zeeman torque (ZT) to influence the spin dynamics \cite{Mondal2016}. Previous numerical studies performed with a damping parameter of $\alpha = 0.02$, demonstrated that in antiferromagnets such as NiO and CoO, the inclusion of the FDT leads to a substantial increase in the amplitude of magnonic oscillations, accompanied by an additional phase shift of $\pi/2$ \cite{Mondal2019PRB}. Recent experiments have provided direct evidence for the existence of the FDT in Bi-doped ferrimagnetic $\mathrm{Gd_{\frac{3}{2}}Yb_{\frac{1}{2}}BiFe_{5}O_{12}}$ ~\cite{Dutta2024,Blank2021THz}. Studies employing ultrafast nonlinear magnetization dynamics have demonstrated that the experimental observations can only be accurately reproduced when the FDT is included alongside the conventional  ZT~\cite{Dutta2025}. Moreover, it has been shown that magnetization reversal is enhanced with the FDT in the same ferrimagnetic iron garnet system~\cite{mukherjee2025ferri}. Despite these advances, both experimental and theoretical explorations of the FDT remain limited, particularly in understanding the role of FDT in antiferromagnetic switching. { A thorough quantitative investigation of FDT-driven dynamics in a wider set of antiferromagnetic materials has not yet been reported in the literature.}

In this letter, we computationally explore the impact of the FDT on ultrafast magnetization switching in antiferromagnetic NiO driven by circularly polarized THz pulses. 
Our results indicate that incorporating the FDT significantly lowers the required THz magnetic field strength for magnetization reversal compared to using only the ZT. For instance, when only the ZT is included, a THz pulse with a temporal width of $\sigma = 4$ ps and a peak magnetic field of 8.15 T fails to switch the magnetization. In contrast, when the FDT terms are included, magnetization reversal occurs under similar conditions. Further, we find that in the absence of the FDT, a much stronger field of 14.25 T is required to achieve magnetization switching. {Moreover, we compute spin-switching diagrams as a function of THz magnetic field strength, THz pulse width, and the Gilbert damping.} These diagrams clearly delineate regions where switching occurs and where it does not, both with and without the inclusion of FDT. Notably, the switching regions expand when FDT effects are incorporated, while the overall diagrams continue to exhibit a periodic pattern of alternating switched and non-switched regions. We have quantified the increase in the switched area resulting from the inclusion of FDT. Finally, we elucidate the switching mechanisms with and without the inclusion of FDT. These mechanisms differ fundamentally from the angular momentum transfer-induced switching observed in ferrimagnets via a transient ferromagnetic state \cite{ostler12}. 

\begin{figure}[H]
    \centering
\includegraphics[scale = 0.4]{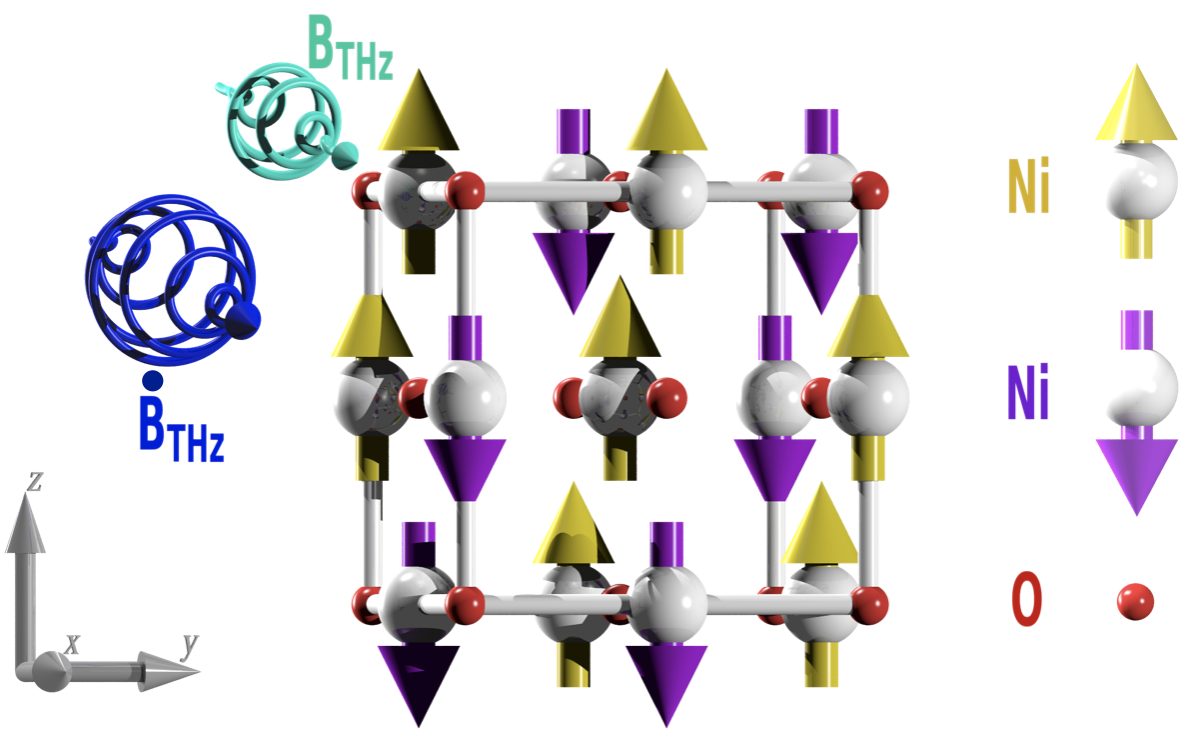}
    \caption{Schematic diagram for circularly polarized THz excitation of antiferromagnetic spins in $\rm NiO$. The upward Ni spins are denoted in golden color, while the downward Ni spins are denoted in purple. {Red small spheres denote the oxygen atoms.}}
    \label{fig:1}
\end{figure}

\section*{Atomistic Spin Dynamics}
To investigate spin dynamics without and with FDT, we have simulated the LLG equation supplemented by the FDT terms as follows \cite{mukherjee2025ferri,Dutta2024,Mondal2016,Blank2021THz}
\begin{multline}
\label{Eq1}
\dot{{\bf S}}_{i} = -\frac{\gamma_i}{1+\alpha_i^2} \,{\bf S}_{i} \times \bigg[
\left({\bf B}_{i}^{\rm eff} -\frac{\alpha_i a^3_i}{\gamma_i\mu_{\rm B}\mu_{\rm 0}}\dot{\textbf{B}}_{\rm THz}\right) \\
+ {{\alpha_i}}\left({\bf S}_i\times
   \left({\bf B}_{i}^{\rm eff} -\frac{\alpha_i a^3_i}{\gamma_i\mu_{\rm B}\mu_{\rm 0}}\dot{\textbf{B}}_{\rm THz}\right)\right)
\bigg].
\end{multline}
The right side of this torque equation contains two terms: the first governs the precession of the magnetic spins, and the second represents the damping torque responsible for energy dissipation. In the above equation, ${\bf S}_i = \frac{\mathbf{\mu}_{s}^{i}}{\vert \mathbf{\mu}_{s}^{i}\vert}$ denotes the normalized spin moment of the $i$th sublattice  and $\mathbf{ \mu}_{s}^{i}$  is the spin moment of the respective sublattices. $\gamma_i$ and $\alpha_i$ represent the gyromagnetic ratio and the phenomenological Gilbert damping of the sublattices, respectively. ${\bf B}_{i}^{\rm eff}$ is the effective field of the system and includes exchange interactions, magnetic anisotropy, and Zeeman field contributions. Such an effective field can be computed through
\begin{align}
\label{Eq2}
\mathbf{{B}}_{i}^{\rm eff} = -\frac{1}{\mu_s^i}\frac{\delta \mathcal{H}}{\delta \mathbf {S}_{i}}.
\end{align}
The other torque term, the FDT, is represented by ${\bf S}_{i} \times \frac{d\textbf{B}_{\rm THz}}{dt}$ in Eq.~(\ref{Eq1}). Notably, the FDT contains both field-like and damping-like components, analogous to the effective magnetic field in the conventional LLG equation. Such a correction to the effective field was first theoretically proposed ~\cite{Mondal2016}, and was shown to depend on a combination of universal constants and material-specific parameters, including the Gilbert damping $\alpha_i$, gyromagnetic ratio $\gamma_i$, and the unit-cell volume per spin $a^3_i$. Such dependence was later experimentally realized recently~\cite{Blank2021THz,Dutta2025}.
We note that the AFMs like NiO and CoO have the same two sublattices oriented in opposite directions, as shown in Fig. \ref{fig:1}. Therefore, the effective field and FDT terms are the same for these two sublattices.

In the above Eq. (\ref{Eq2}), $\mathcal{H}$ represents the spin Hamiltonian of the system representing the total energy. Such an atomistic spin Hamiltonian can be written as
\begin{align}
\label{Eq3}
    \mathcal{H} &= \sum_{i<j} J_{ij} \mathbf{S}_i \cdot \mathbf{S}_j
    - \sum_{i} (d_x {S_{i}^{x}}^{2} + d_y {S_{i}^{y}}^{2})
    \nonumber\\
    & - \sum_{i} {\mu}_s^{i} \mathbf{B}_{\rm THz}(t) \cdot \mathbf{S}_i\,.
\end{align}
The first term on the right side represents the isotropic Heisenberg exchange energy, where $J_{ij}$ is the magnetic exchange energy.  We take into account interaction with the nearest neighbors (NN) and the next nearest neighbors (NNN) as $J_{ij}^{\rm NN}$ and $J_{ij}^{\rm NNN}$, respectively. The second term represents the magnetic anisotropy energy, where $d_x$ and $d_y$ are the in-plane biaxial anisotropy energies.The third term represents the Zeeman energy arising from the applied time-dependent THz magnetic pulse.
\begin{figure*}
    \centering
\includegraphics[width=0.45\textwidth]{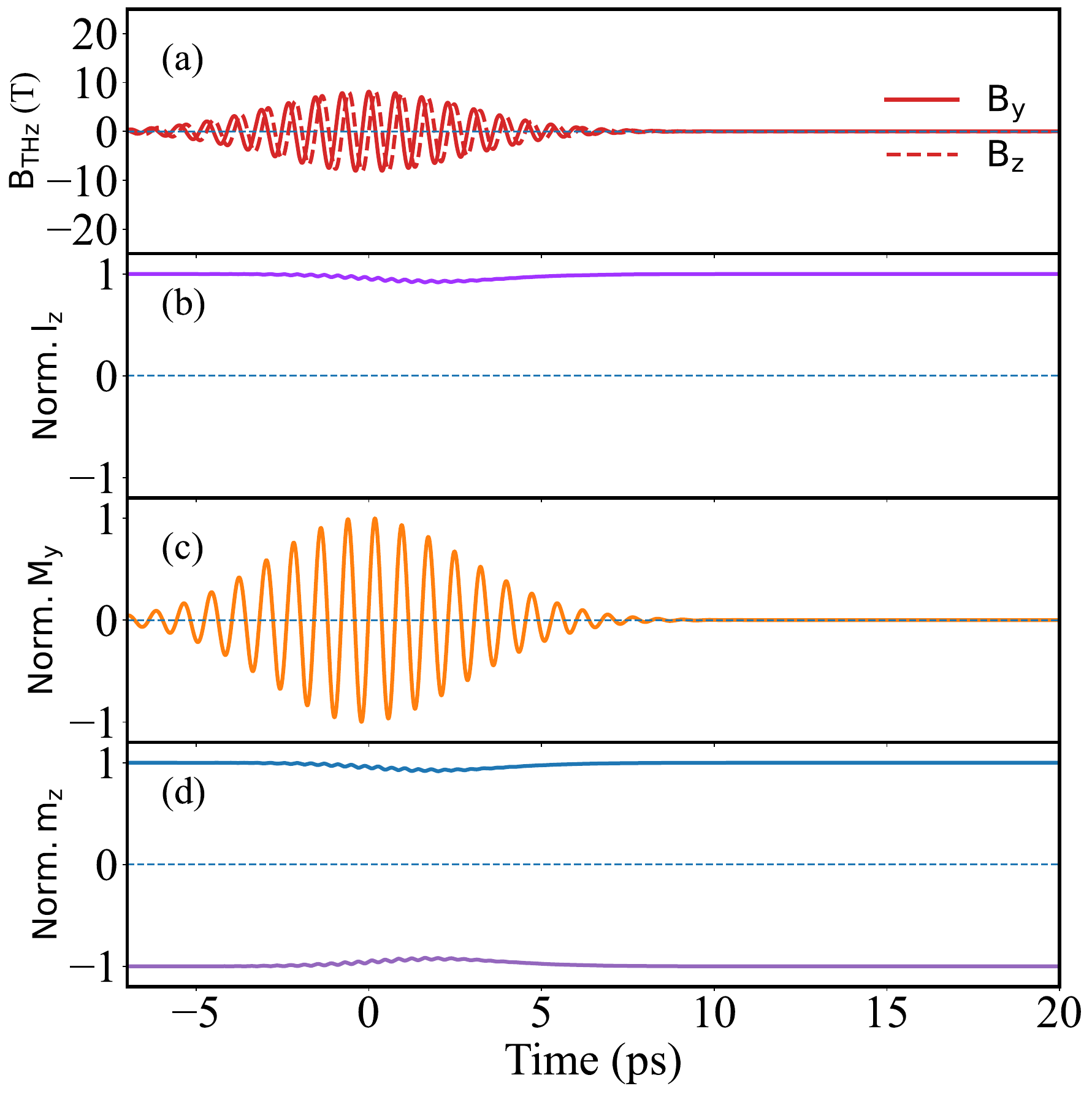}
\includegraphics[width=0.45\textwidth]{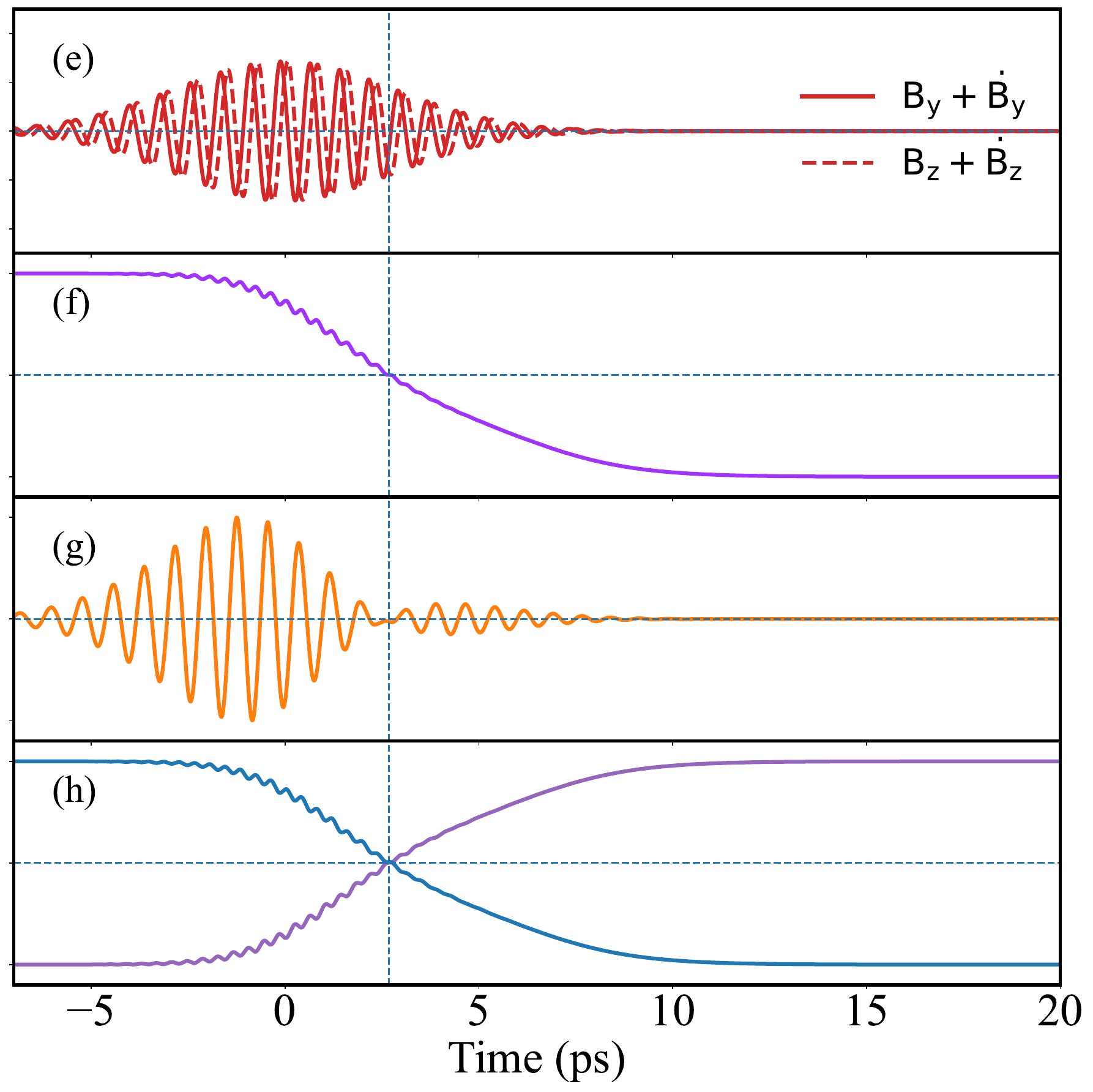} \caption{Simulation results of the magnetization dynamics without FDT (Left Panel) and with FDT (Right Panel) for an incident THz pulse having ${\rm B_0}= 8.15$\,T, $\sigma = 4$\,ps and $\tau = 130$\,ps. (a) and (e) show the $y$- and $z$-components of the circularly-polarized THz pulse, (b) and (f) show the $z$-component of the Néel vector, (c) and (g) show the $y$-component of the total magnetization, while (d) and (h) show magnetization dynamics of the individual sublattices.}
    \label{fig:2}
\end{figure*}

We perform atomistic spin simulations at $T = 0$ K for NiO, considering { one Ni atom for each sublattice} with antiferromagnetically coupled Ni sublattices excited by a circularly polarized THz pulse travelling along the $x$-direction. The THz pulse frequency is set to ${\rm f_0} = 1.3$ THz, which has been reported as the resonance frequency of $\rm NiO$ \cite{Hutchings1972,Archer2011}. The THz pulse is characterized by its field amplitude $B_0$, pulse width $\sigma$, and chirp time $\tau$, as detailed in the Supplemental Material S1~\cite{SuppleMater}. Further, we consider the NN ferromagnetic coupling and a strong NNN antiferromagnetic coupling along with the biaxial anisotropy energies reported in Hutchings {\it et al.} \cite{Hutchings1972}. The simulation setup has been schematically shown in Fig. \ref{fig:1}. Importantly, the Ni spins experience torque contributions from both the THz field ${\bf B}_{\rm THz}(t)$ and its time derivative ${\dot{\bf B}}_{\rm THz}(t)$. Denoting $\mathbf{m}$ as the magnetization of the individual sublattices, we define the Ne\'el vector as $\mathbf{l} = { \left(\mathbf{m}_{\rm Ni_{\uparrow}} - \mathbf{m}_{\rm Ni_{\downarrow}}\right)/2} $, where $\mathbf{l} = \rm l_x \hat{x} + \rm l_y \hat{y} + \rm l_z \hat{z}$ and the total magnetization as $\mathbf{M} = { \left(\mathbf{m}_{\rm Ni_{\uparrow}} + \mathbf{m}_{\rm Ni_{\downarrow}}\right)/2} $, where $\mathbf{M} = \rm M_x \hat{x} + \rm M_y \hat{y} + \rm M_z \hat{z}$. Further details of the NiO material parameters are provided in Supplemental Material S2~\cite{SuppleMater}.   

\section*{Results and Discussions}

To elucidate the influence of the FDT, we calculate the N\'eel vector and total magnetization dynamics, 
together with the switching behavior of NiO in both cases, without and with FDT. The corresponding results are displayed in Fig.~\ref{fig:2}. A comparison of the THz fields shown in Fig.~\ref{fig:2}(a) and Fig.~\ref{fig:2}(e) clearly demonstrates that incorporating the FDT leads to a substantial increase in the effective field amplitude. 
 The normalized $\rm l_z$ components shown in Figs.~\ref{fig:2}(b) and~\ref{fig:2}(f) highlight the contrasting behaviors without and with the FDT, respectively. In the absence of FDT [Fig.~\ref{fig:2}(b)], the weak torque induces oscillatory magnetization dynamics, but the N\'eel vector ultimately relaxes back to its initial state. When the FDT is included [Fig.~\ref{fig:2}(f)], however, the enhanced torque drives a complete switching of the N\'eel vector.   
In the ground state, the NiO magnetization points along the $z$-axis, so the $\mathbf{B}^z_{\rm THz}$ component of the circularly polarized THz pulse initially has no effect on the dynamics because its precessional torque vanishes. The $\mathbf{B}^y_{\rm THz}$ component, however, exerts a finite torque, tilting the spins away from equilibrium. This tilt activates the coupling to $\mathbf{B}^z_{\rm THz}$, which then begins to contribute to the spin dynamics.
\begin{figure}[tbh!]
    \centering
\includegraphics[width=1.05\linewidth]{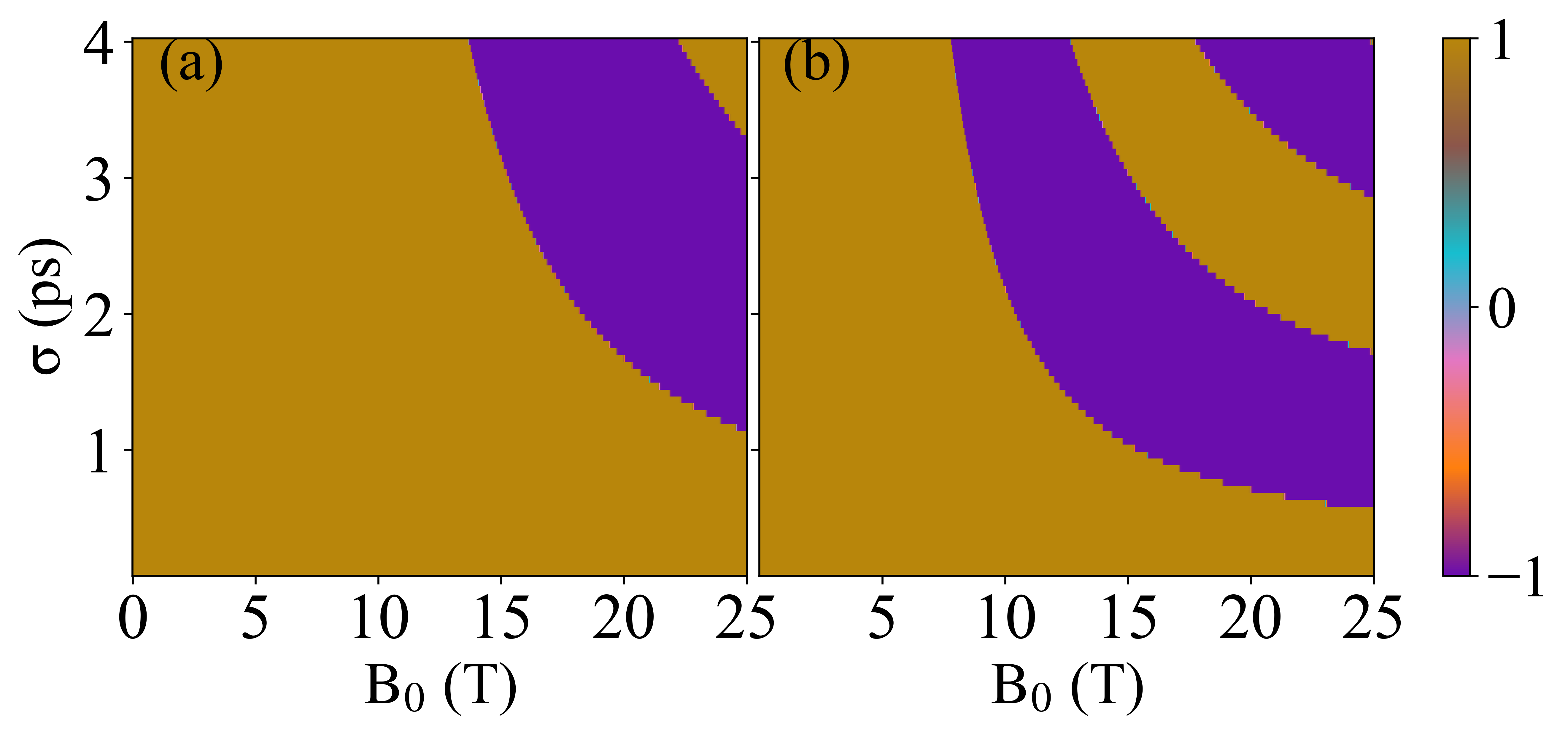}
    \vspace{-15pt}
    \caption{Normalized ${\rm l_z}$ after THz excitation of NiO as a function of THz field amplitude $B_0$ and pulse width $\sigma$. Golden regions (${\rm l_z} = 1$) indicate no reversal, while purple regions (${\rm l_z} = -1$) denote magnetization switching. (a) Without FDT and (b) with FDT.}
    \label{fig:3}
\end{figure}

With the circularly polarized THz pulse applied along $x$ and the initial magnetization along $z$, the resulting effective torque acts along $y$. Initially, the spins are collinear along $z$, but the pulse induces spin cantations, as reflected in the dynamics of the normalized total magnetization $\rm M_y$ in Figs.~\ref{fig:2}(c) and~\ref{fig:2}(g). As shown in Fig.~\ref{fig:2}(c), the normalized $\rm M_y$ component starts from zero and, after a few smooth precessions, relaxes back to its initial value. 
On the contrary, when the FDT is included [Fig.~\ref{fig:2}(g)], the precession of $\rm M_y$ becomes irregular and is disrupted precisely when the switching of $\rm l_z$ occurs. At this point, $\rm M_y$ momentarily reaches almost zero, indicating that the net magnetization along $y$ vanishes and that the spins of the two Ni sublattices are oppositely aligned along this direction. This corresponds to a transient reorientation of the spins from the $z$-direction toward the $y$-direction. Since $\rm l_z$ also becomes zero at the same moment and reverses afterward, we conclude that the switching is mediated by a $90^{\circ}$ spin reorientation through precessional motion. The individual sublattice magnetizations, shown in Figs.~\ref{fig:2}(d) and~\ref{fig:2}(h) for cases without and with FDT, clearly demonstrate the occurrence of magnetization switching. Such a switching mechanism is fundamentally different than the ferrimagnetic switching mediated by the transfer of angular momentum \cite{ostler12}.

\begin{figure}[tbh!]
    \centering
    \includegraphics[width=1.2\linewidth]{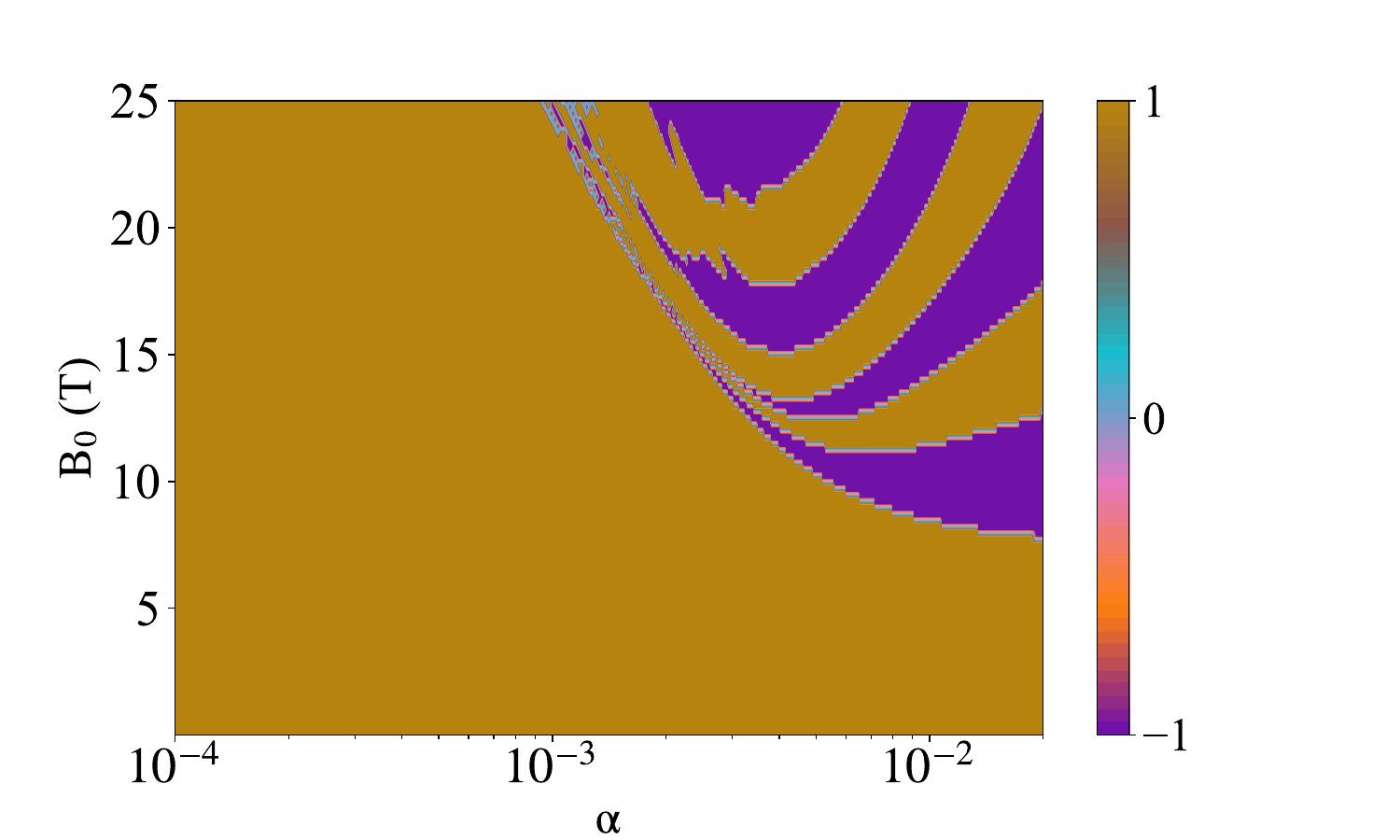}
\caption{Normalized ${\rm l_z}$ after THz excitation of NiO as a function of THz field amplitude $B_0$ and Gilbert damping $\alpha$ with FDT. Golden regions (${\rm l_z} = 1$) indicate no reversal, while purple regions (${\rm l_z} = -1$) denote switching.}
    \label{fig:4}
\end{figure}
In AFMs, the magnetization switching does not occur near the peak of the THz pulse but rather after the pulse maximum has passed. This behavior contrasts with the FDT-mediated magnetization reversal observed in ferrimagnets \cite{mukherjee2025ferri}. Nonetheless, the magnetization reversal happens at a lower THz magnetic field with the inclusion of the FDT.

We further investigate how the required THz field amplitude varies with the pulse temporal width, with and without FDT. Figure~\ref{fig:3} presents the norm. {$\rm l_z$} as a function of the field amplitude $B_0$ and pulse width $\sigma$. The golden regions denote relaxation of the N\'eel vector to its initial state ($\rm l_z = 1$), whereas the purple regions mark complete reversal ($\rm l_z = -1$).
Figure~\ref{fig:3}(a) shows the switching and non-switching regions without FDT, while Fig.~\ref{fig:3}(b) presents the same with FDT.  
We observe that for larger $\sigma$, a smaller field amplitude is sufficient to reverse the magnetization. With the inclusion of FDT at $\sigma = 4$ ps, magnetization switching occurs at a field amplitude of 8.15 T. In contrast, without the inclusion of FDT [see Fig.~\ref{fig:3}(a)], magnetization switching does not occur at this field; a significantly stronger THz field of 14.25 T is necessary to induce reversal, reinforcing our earlier discussions. Such switching results obtained without the inclusion of FDT are discussed in Supplemental Material S3~\cite{SuppleMater}. Additionally, a periodic pattern of switched and non-switched regions emerges in both cases, consistent with earlier investigations \cite{Sonke2012}. This periodic behavior is more pronounced and coherent than that observed in ferrimagnets~\cite{mukherjee2025ferri}. While the transition from non-switching to switching regions in the AFM is sharp, the corresponding change in ferrimagnets is gradual \cite{mukherjee2025ferri}, highlighting the presence of demagnetized states in ferrimagnets, which are absent in AFMs. Nonetheless, the switching area increases by $\sim$ 55\% when FDT is included (see Supplemental Material S4)~\cite{SuppleMater}.

So far, the switching regions have been analyzed in terms of variations in the THz pulse parameters e.g., $B_0$ and $\sigma$. To further investigate the material-specific properties and the corresponding FDT effect, we additionally compute the switching regions as a function of THz magnetic field $B_0$ and Gilbert damping $\alpha$. The corresponding results are plotted in Fig. \ref{fig:4} with FDT.  In Fig.~\ref{fig:4}, we consider $\alpha$ in the range from $10^{-4}$ to $2 \times 10^{-2}$, which spans the two limiting cases of damping relevant to the system under study. At one end, the damping value corresponds to that of NiO reported experimentally~\cite{Kampfrath2011} as $(2.1 \pm 0.1) \times 10^{-4}$, however, we use a higher value $\alpha = 0.02$ in our simulations so far \cite{Mondal2019PRB}. We observe that for sufficiently small values of $\alpha$, magnetization switching does not occur even at higher field amplitudes, despite the inclusion of FDT. In contrast, for larger $\alpha$, magnetization reversal is observed even at comparatively lower field amplitudes. These findings highlight the strong dependence of the FDT-induced switching effects on the material-dependent Gilbert damping, as suggested by Eq.~(\ref{Eq1}). Similar to Fig.~\ref{fig:3}, the periodic alternation between switching and non-switching regimes is clearly visible. The switching behavior as a function of THz pulse width $\sigma$ and Gilbert damping $\alpha$ is discussed in Supplemental Material S5~\cite{SuppleMater}.

\begin{figure}[tbh!]
    \centering
\includegraphics[width=1\linewidth]{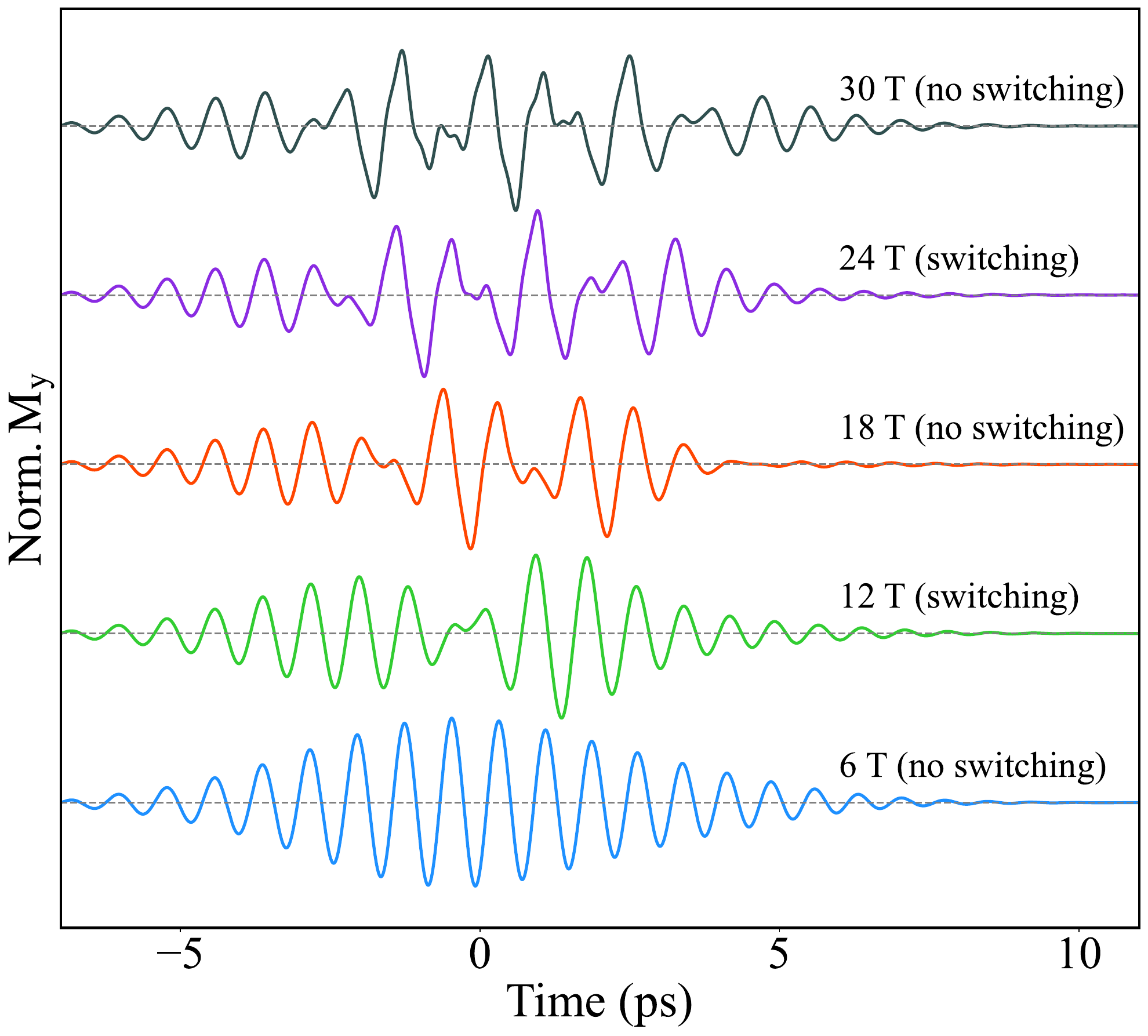}
    \caption{The dynamics of total magnetization ${\bf M}$ along the effective torque direction exerted by the THz pulse and the FDT.}
    \label{fig:5}
\end{figure}

The emergence of switching and non-switching regimes can be further understood by computing the norm. ${\rm M_y}$ component as a function of time for several THz field amplitudes, with the Gilbert damping fixed at $\alpha = 0.02$.  
As the field amplitude increases, both the precessional and damping torques in the LLG equation become stronger; the interplay between these competing torques together with FDT enhances the overall precession. The resulting increase in precession, together with the exchange coupling between the two sublattices, governs whether a 90$^{\rm o}$ reorientation of the magnetization can occur, thereby determining the switching of the N\'eel vector relative to its initial direction. In Fig.~\ref{fig:5}, for ${\rm B_0} = 6~\mathrm{T}$, the norm. ${\rm M_y}$ component oscillates around zero without exhibiting any phase change. Such behavior indicates the absence of switching as the THz pulse passes through. However, for ${\rm B_0} = 12~\mathrm{T}$, the norm. ${\rm M_y}$ component oscillates around zero and reaches a value near zero once, indicating a phase change. Such a change in phase induces a single reorientation of the spins from the equilibrium $z$-direction toward the $y$-direction. This reorientation mediates the switching of the Néel vector from $+z$ to $-z$. For ${\rm B_0} = 18~\mathrm{T}$, the normalized ${\rm M_y}$ reaches a value near zero twice, indicating two successive spin reorientations. Following the first reorientation, the Néel vector switches from $+z$ to $-z$, and after the second reorientation, it switches back from $-z$ to $+z$. Overall, the total magnetization dynamics determine whether switching occurs. The observed switching and non-switching regimes can be understood in terms of phase changes in the total magnetization dynamics. Specifically, the absence of phase change or the occurrence of an even number of phase changes corresponds to a non-switching regime, whereas an odd number of phase changes indicates a switching regime. { So far, we have discussed the magnetization switching regimes for THz excitation at the resonant frequency of NiO. When exploring off-resonant THz excitation, the FDT significantly enlarges the switching area compared to the case without the FDT, as discussed in Supplemental Material S6~\cite{SuppleMater}.}    

We now compare the results obtained for ferrimagnets and AFMs. In contrast to ferrimagnetic systems~\cite{mukherjee2025ferri}, the inclusion of FDT in AFMs does not generate an asymmetric torque between the two sublattices, since the opposing magnetic moments are equal in magnitude and both originate from Ni atoms. The mechanism for ferrimagnetic switching relies on the torque transfer between two unequal sublattices, while the AFM switching is mediated via the phase change in the total magnetization dynamics.  Nonetheless, the ZT+FDT enhances the total switching area compared to the non-switching area when only ZT is considered. It is worth noting that a larger THz pulse width is required to induce switching in AFMs, whereas a relatively smaller pulse width is sufficient for ferrimagnets. This difference arises from the much stronger antiferromagnetic exchange energy in AFMs compared to ferrimagnets. Similar large pulse widths have also been employed for AFM switching in the absence of FDT~\cite{Sonke2012}. 

{ There are two key challenges that impede the observation of FDT-driven magnetization switching in antiferromagnets. The first is the presence of enhanced Gilbert damping, and the second is the requirement of a circularly polarized THz pulse with a sufficiently strong THz magnetic field.} 
In our simulations, a relatively high Gilbert damping $\alpha$ was used, since FDT effects become more prominent in materials with larger damping~\cite{Mondal2019PRB,Blank2021THz}. However, the experimentally reported damping for NiO is comparatively small. For instance, monocrystalline and polycrystalline NiO exhibit Gilbert damping values of $\alpha = (5.0 \pm 0.4) \times 10^{-4}$ and $\alpha = (7.4 \pm 0.4) \times 10^{-4}$, respectively, arising from intrinsic and extrinsic effects~\cite{Moriyama2019}. { The extrinsic effects include the presence of impurity or defects.} Nevertheless, the damping can be enhanced experimentally by engineering heterostructures of NiO with heavy metals possessing strong spin–orbit coupling ~\cite{barati14,Hidetoshi2025}. In such systems, interfacial spin pumping and spin–orbit torque can significantly increase the effective damping, thereby amplifying the FDT-induced spin torques that promote magnetization switching. Furthermore, temperature-dependent measurements or impurity engineering can also modify the effective damping and enable experimental observation of the predicted switching dynamics~\cite{Kumar2017,Haspot2022}. { The temperature-dependence change in Gilbert damping is maximised for thin films~\cite{Zhang2025PRM}}. These considerations suggest that tuning the damping parameter through material design or interface engineering offers a practical route to realizing FDT-driven ultrafast switching in AFMs. { To address the second issue, strong circularly polarized THz pulses can be generated via tilted pulse-front THz in LiNbO$_3$, plasma oscillators, or Mach-Zehnder methods ~\cite{Li_2024,Ueda2025,Wu2008}. Linearly polarized multi-cycle THz pulses having strength about 30 T can also be converted to circular polarization ~\cite{Junginger2010,Hoffmann2011}. }

\section*{Conclusions}

To conclude, we have investigated the role of the FDT in THz-induced magnetization dynamics and switching of antiferromagnetic NiO. Our simulations reveal distinct switching and non-switching regimes that strongly depend on the Gilbert damping parameter, the THz field amplitude, and the THz pulse width. The inclusion of FDT is found to significantly enhance the switching efficiency, reducing the required THz field amplitude for deterministic magnetization reversal by approximately two-fold. { Other intrinsic torques, such as the optical spin-orbit torque, have been discussed in the literature, although their role in magnetization switching remains unexplored. The inertial spin torque, effective mainly in low-damping materials through nutational switching, may also reduce the required field amplitude~\cite{Winter2022}. However, both mechanisms are fundamentally distinct from the high-damping-favored FDT-driven precessional switching studied here.}

\section*{acknowledgments}
We thank Peter M. Oppeneer, Shovon Pal, and Arpita Dutta for fruitful discussions. 
The authors acknowledge funding support from the SERB-SRG via Project No. SRG/2023/000612 and the faculty research scheme at IIT (ISM) Dhanbad, India, under Project No. FRS(196)/2023-2024/PHYSICS.

  
\section*{DATA AVAILABILITY STATEMENT}
The data that support the findings of this study are available from the corresponding author upon reasonable request.


%
\providecommand{\noopsort}[1]{}\providecommand{\singleletter}[1]{#1}%
\end{document}